\documentstyle[11pt,newpasp,psfig,twoside]{article}
\index{instructions}
\index{guidelines}

\def\edcomment#1{\iffalse\marginpar{\raggedright\sl#1\/}\else\relax\fi}
\reversemarginpar

\begin{document}
\title{Observational manifestations of young neutron stars: 
Spin-powered pulsars}
\author{Michael~Kramer}
\affil{University of Manchester, Jodrell Bank Observatory, UK}

\begin{abstract}
The largest number of known young neutron stars are observed as
spin-powered pulsars. While the majority of those are detected at radio
frequencies, an increasing number can be studied at other parts of the
electromagnetic spectrum as well. The Crab pulsar is the prototype of
a young pulsar which can be observed from radio to gamma-ray
frequencies, providing a red thread of discussion during a tour through
the pulsar properties observed across the electromagnetic spectrum.
The basic observational features of pulsar emission are presented,
preparing the ground for more detailed reviews given in these
proceedings. Here, particular attention will be paid to those emission
features which may provide a link between the radio and high-energy
emission processes.
\end{abstract}

\section{Introduction}

It is an impossible task to summarize the diversity of pulsar
phenomena observed across the electromagnetic spectrum on a few
pages. Clearly, the wealth of information that can be obtained outside
the classical radio window today justifies some re-definition of the
word ``pulsar''. Rather than being characterised by appearing as a
pulsating {\em radio} sources, a new definition of {\em pulsar} should
be that it {\em emits radiation that is pulsed due to rotation and
powered by rotational energy}. This definition also encompasses X-ray
pulsars that are clearly powered by the loss of rotational energy but
which have not been detected at radio frequencies. This may be due to
a misaligned radio beam or due to a very low radio luminosity.  The
discovery of an increasing number of very weak radio pulsars
coincident with X-ray point sources by Camilo and co-workers (see
these proceedings) clearly shows that the expression ``radio-quiet''
cannot be used without the discussion of a corresponding flux density
limit. Bearing this in mind, we will use ``radio-quiet'' in the
context of sources where pulsations have been detected at high
energies but not (yet) at radio frequencies, assuming that there is no
fundamental difference in the physics of these objects. Also,
concentrating on young pulsars, we will neglect the whole population
of millisecond pulsars which appear, however, to function under the
same underlying principles as young pulsars (Kramer et al.~1998).
There is not sufficient room to discuss these principles in a detailed
theoretical framework, but observational implications for a working
theory will be pointed out. Young, slowly-rotating pulsars known
as ``Magnetars'' (SGRs/AXPs) are discussed by Kaspi in these
proceedings.

\vspace{-0.3cm}

\section{The population of young pulsars}

\begin{figure}[h!]
\centerline{\psfig{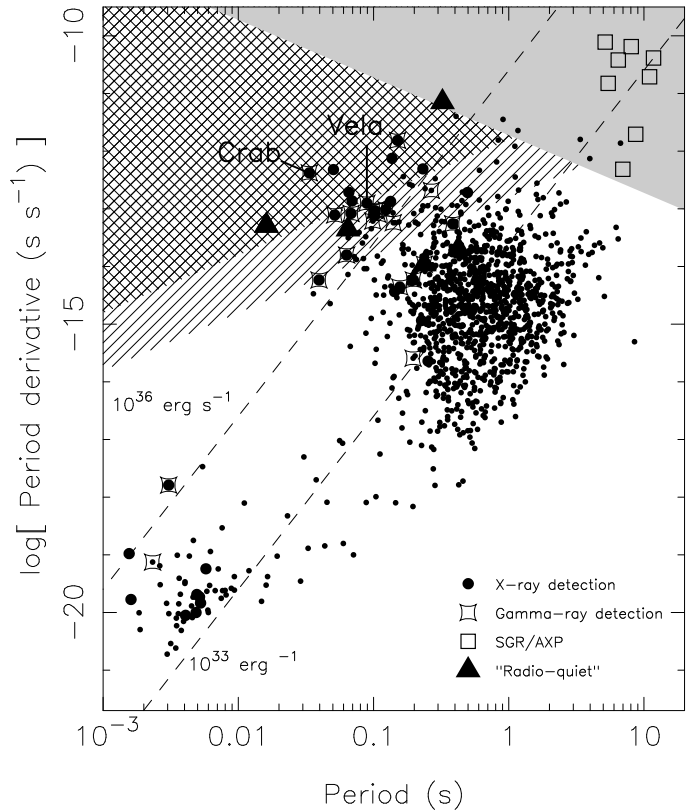}}

\caption{\label{fig:ppdot} $P-\dot{P}$--diagram of the currently
known population of pulsars. The hatched area contains pulsars
with spin-down ages of less than 100 kyr, whilst the cross-hatched
area marks young pulsars with an age less than 10 kyr. Pulsars
located in the grey area exhibit surface magnetic fields,
$B_S = 3.2\times 10^{19}\sqrt{P\dot{P}}$ Gauss, above the
quantum critical field. Lines of constant spin-down luminosity
of $\dot{E}=10^{33}$ erg s$^{-1}$ and $\dot{E}=10^{36}$ erg s$^{-1}$
are shown. Pulsars marked as large filled circles have been detected
at X-ray frequencies, and pulsars with $\gamma$-ray detections are
shown by sharply-edged squares. Normal squares indicate neutron stars
identified as SGRs or AXPs. Spin-powered pulsars with not (yet?)
visible as normal radio pulsars are shown as filled triangles.
}
\end{figure}

The majority of spin-powered pulsars are still discovered and studied
at radio frequencies. They dominate by far the number of about 1600
pulsars currently known, of which most are shown in the
$P-\dot{P}$--diagram presented in Figure~\ref{fig:ppdot}.  We will
define ``young pulsars'' as those with a characteristic age,
$\tau=P/2\dot{P}$, of less than 100 kyr. About 80 of these are
currently known, and they are located in the upper left area of
Fig.~\ref{fig:ppdot}.  Specifically, pulsars with characteristic ages
of less than 10 kyr appear in the cross-hatched area, whilst pulsars
with ages between 10 and 100 kyr are located in the hatched area. The
latter pulsars are often compared to the Vela pulsar if they match its
spin-down luminosity, i.e.~$\dot{E}\ga 10^{36}\;$erg~s$^{-1}$. A
corresponding line of constant $\dot{E}$ is shown together with a line
for $\dot{E}=10^{33}$ erg s$^{-1}$. About 26 Vela-like pulsars are
currently known (Kramer et al.~2003a). In comparison, there are only 6
Crab-like pulsars, including the fastest rotating young but
radio-quiet pulsar PSR J0537$-$6910 with a period of only 16 ms, the
``Crab-twin'' PSR B0540$-$69, PSR J1124$-$5916, and PSR B1509$-58$. In
the same group we also find PSR J1119$-$6137 with the youngest age and
highest $\dot{P}$ for any radio pulsar whilst the overall records are
currently set by the radio-quiet PSR J1846$-$0258 in Kes75 with a
period of 323 ms and characteristic age of only 722 yr (Gotthelf et
al.~2000). This raises the question about the birth periods of pulsars
which have currently been estimated for about 9 sources
(e.g.~Migliazzo et al.~2002), covering one order of magnitude from
less than 14 ms for PSR J0537$-$6910, over 19 ms for the Crab pulsar
to 140 ms for PSR J0538+2817 (Migliazzo et al.~2002, Kramer et
al.~2003b). For some of these estimates, a measured
braking index was needed. In total, 5 pulsars have a measured braking
index that seems to reflect a physical braking process (rather than
timing noise) with values ranging from $n=1.4$ to $n=2.9$ 
(e.g.~Zhang et al.~2001).

\vspace{-0.3cm}

\section{Radio properties}
\label{radio}

The radio properties of pulsars are discussed in detail by Gupta
(these proceedings). Therefore, we do not attempt to present or
discuss the rich assortment of radio phenomena, but shortly review
those which we consider to be important in context of the bigger
picture and their connection to high energy emission. In comparison to
high energies, the energy output in the radio is tiny (see
Section~\ref{relation}), with a median luminosity observed at 1400 MHz
of $L_{1400}=25\pm 5$ mJy kpc$^2$. This number is biased by distant
luminous sources, so that $L_{1400}=3\pm 1$ mJy kpc$^2$ computed for
pulsars within a 3 kpc distance is a more representative number. There
is a slight trend for older pulsars to be less luminous but this is
mostly suggested by the millisecond pulsars which are indeed less
luminous and less efficient radio emitters (Kramer et al.~1998). The
flux density spectrum, $S\propto \nu^\alpha$, is steep with a mean
spectral index of $\alpha \sim-1.7$. No age dependence is found
despite earlier reports to the contrary (Maron et al.~2000). The
highest radio frequency that pulsars have been detected at is 87 GHz
(Morris et al.~1997) and there are indications for a spectral turn-up
at mm-wavelengths (Kramer et al.~1996). These observations are
interesting due to a suggested ``radius-to-frequency mapping'' (RFM),
according to which high frequency radio emission originates from
closer to the neutron star surface than low frequency emission (Cordes
1978).  The model is motivated by the observation that pulse profiles
become narrower at high frequencies. However, it seems difficult to
decide whether the derived magnetospheric altitudes correspond to the
heights where the coherent, usually highly polarized emission is {\em
created} or where it {\em escapes} the magnetosphere after it had been
created further down and propagated to the escape radius, possibly in
certain wave modes (see contributions by Gupta, Petrova and
Karastergiou et al.).  In any case, there seems to be no doubt that
the classical radio emission originates from
a few hundred km or so above the pulsar
surface (e.g.~Kramer et al.~1997), and that the radiating plasma
originates from the polar cap region.  The interesting question is
thereby how this plasma is related to that creating high energy
emission, for which competing models place the origin near the polar
cap as well (e.g.~Daugherty \& Harding 1996) or further out in the
magnetosphere in so-called ``outer gaps'' (e.g.~Cheng et al.~1986).
Clues are available from {\em giant pulses} observed for a handful of
pulsars at radio frequencies. Giant pulses are individual pulses with
flux densities exceeding the mean value by a factor of 10 to 100 or
more. Recently published results for the Crab pulsar by Hankins et
al.~(2003) reveal fine-structure on nanosecond timescales with
brightness temperatures reaching $10^{37}$ K.  Giant pulses tend to be
aligned with the high energy pulses rather than with the radio 
profile (see Section~\ref{giants}).  This suggests a common origin,
indicating that some observed radio emission could be in fact a
by-product of the high energy radiation process.  This could explain
the highly unusual ``High-Frequency Components'', seen to emerge at
some odd pulse phases in the Crab profile at a few GHz (Moffett \&
Hankins 1996), as the results of such possible by-products and
geometrical RFM-like effects. This likely possibility underlines our
need to take geometry into account if we want to understand both the
observed radio as well as the high energy emission.

\vspace{-0.3cm}

\section{Optical properties}
\label{optical}

Currently, only five pulsars are reported to show pulsed optical
emission (see contribution by Mignani or Shearer \& Golden 2002 for a
review). The prime example is the Crab pulsar which is strong enough
to allow extensive studies of its single pulses and their polarization
properties, revealing interesting differences between the peak
emission of the prominent double-peaked profile and the bridge and
non-zero off-pulse emission (e.g.~Romani et al.~2001, Kanbach et
al.~2003). The similarity between the optical, X-ray and $\gamma$-ray
profiles suggests that the emission is created by the same incoherent
non-thermal radiation process, presumably at the same location.  The
polarization measurements and the applications of the rotating-vector
model (Radhakrishnan \& Cooke~1969) thereby allow us to study the
geometry of the corresponding emission region in a way that is not yet
possible at X-rays.  Such
studies are not possible for the other weaker
pulsars with detected pulsed
emission, i.e.~Vela, Geminga and PSRs B0656+14 and B0540$-$69.
Unpulsed emission has been detected for four pulsars, including PSR
B1055$-$52 which together with the Crab, Vela, Geminga and possibly
PSR B0656+14 is also detected as a gamma-ray pulsar
(e.g.~Thompson 2001), again suggesting that the optical emission is
closely related to the X-ray and $\gamma$-ray emission. For a
discussion of the involved luminosities it is important to note that
for both PSRs B1055$-$52 and B0656+14 the most recent distance
estimates dropped by more than a factor of two 
(see Kramer et al.~2003a, Brisken et al.~2003).

\vspace{-0.3cm}

\section{X-ray properties}
\label{xray}

On one hand, X-ray observations promise to offer a more direct insight
into the magnetospheric plasma densities and processes when compared
to the view provided by radio observations that is somewhat occluded
by the (unknown) physics responsible for the coherence of the radio
emission. On the other hand, in many pulsars the observed X-ray
emission is due to a mixture of thermal and non-thermal processes
which cannot always be discriminated by the available
data. Non-thermal, magnetospheric emission can be created by an
accelerated relativistic plasma and should be highly pulsed with a
power-law spectrum ranging from optical to $\gamma$-ray
frequencies. Thermal emission may originate from the stellar surface
and may still show low-amplitude modulation due to the rotating hot
polar cap. In this case, we would expect a blackbody spectrum that is
modified by a possible atmosphere of the neutron star, ranging from
optical to soft X-ray frequencies. In addition, matters may be
complicated due to unresolved emission from pulsar-driven synchrotron
nebulae or pulsar winds interacting with dense ambient
media. Neglecting X-ray detections of millisecond pulsars, pulsed
emission has been observed for 15 radio pulsars while 9 pulsars have
only been seen as unpulsed X-ray sources (Fig.~\ref{fig:ppdot}; see
Becker, these proceedings, and Becker \& Aschenbach~2002 for a
review).  In addition, Geminga and PSRs J0537$-$6910, J1209$-$51/52,
J1811$-$1925 and J1846$-$0258 are clearly spin-powered pulsars at
X-rays but have no corresponding radio detection, yet.  An account of
recent deep searches in X-ray sources is given by Camilo (these
proceedings), which lead, for instance, to the discovery of radio
pulses from the 66-ms pulsar J0205+6449 in SNR 3C58.  This pulsar
would supersede the Crab as the youngest radio pulsar if 3C58 is
indeed associated with the historical supernova SN1181 (Murray et
al.~2002). In general, it is instructive to group the X-ray detected
pulsars into Crab-like sources, Vela-like sources and middle-aged
pulsars. The Crab-like pulsars include PSRs B0540$-$69, J0537$-$6909,
B1509$-$58, J0205+6449 and J1617$-$5055, all of which show strong
features of magnetospheric emission and doubled-peaked profiles which
are, if detected, aligned with similar optical profiles. For Vela-like
pulsars only Vela itself has been detected as a faint optical
source. Their X-ray spectra do not represent simple power-laws whilst
the pulse profiles are more complicated and typically misaligned with
radio or $\gamma$-ray pulses. Finally, the middle-aged pulsars,
i.e.~Geminga, PSR B1055$-$52 and PSR B0656+14, show a mixture of
thermal and non-thermal emission with energy-dependent
profiles. The non-thermal emission is thought to originate either from
the polar cap region (e.g.~Daugherty \& Harding~1996, Harding
et al.~2002) or outer gaps (e.g.~Cheng et al.~1986, Romani~1996).

\vspace{-0.3cm}

\section{$\gamma$-ray properties}
\label{gamma}

The $\gamma$-ray emission of pulsars is highly pulsed, beamed and,
obviously, of non-thermal origin.  There are 7 ``classical''
$\gamma$-ray detections, i.e.~Crab, Vela, Geminga and PSRs B1055+53,
B1509$-$58, B1706$-$44 and B1951+32 (see e.g.~Thompson 2001 or Kanbach
2002 for recent reviews). Their typically double-peaked profiles
change their appearances towards higher energies due to a
phase-dependent spectral hardness, suggesting, again, that geometry
plays an important role. Similar studies are not possible for other,
much weaker pulsars for which a detection has been suggested. Here,
significance levels are much lower, and often their detection is based
on the positional coincidence of a known radio pulsar with an
unidentified $\gamma$-ray
point source.  A viable pulsar candidate should have properties
consistent with those of the classical $\gamma$-ray pulsars,
shown to be steady $\gamma$-ray sources (e.g.~McLaughlin et
al.~1996).  While the beaming fraction is uncertain, the $\gamma$-ray
luminosity seems to be a few per cent of the spin-down
luminosity. Until recently, this $\gamma$-ray efficiency was believed
to reach much larger values, up to 20\%. However, this value was set
by PSR B1055$-$52 for which the updated distance estimate (see
Section~\ref{optical}) now results in a decreased efficiency of only
4\%. A recent detailed discussion of the proposed
associations was presented by Kramer et al.~(2003a)
who estimated that about $19\pm
6$ of the associations are genuine.  Ultimately, it will need
instruments like GLAST to verify most of these candidates. By
extending the observed $\gamma$-ray spectrum to a few tens GeV, GLAST
may then also be able to distinguish between the outer gap and polar
cap model. The polar cap model interprets the observed emission as
inverse Compton scattering of upward polar cap cascades and expects a
cut-off of the spectrum due to $\gamma$-B-field absorption.  In
contrast, the spectrum expected for the outer gap particles is
curvature radiation limited and should extend to higher energies
(e.g.~Thompson 2001).

\begin{figure}[h!]
\centerline{
\begin{tabular}{cc}
\psfig{file=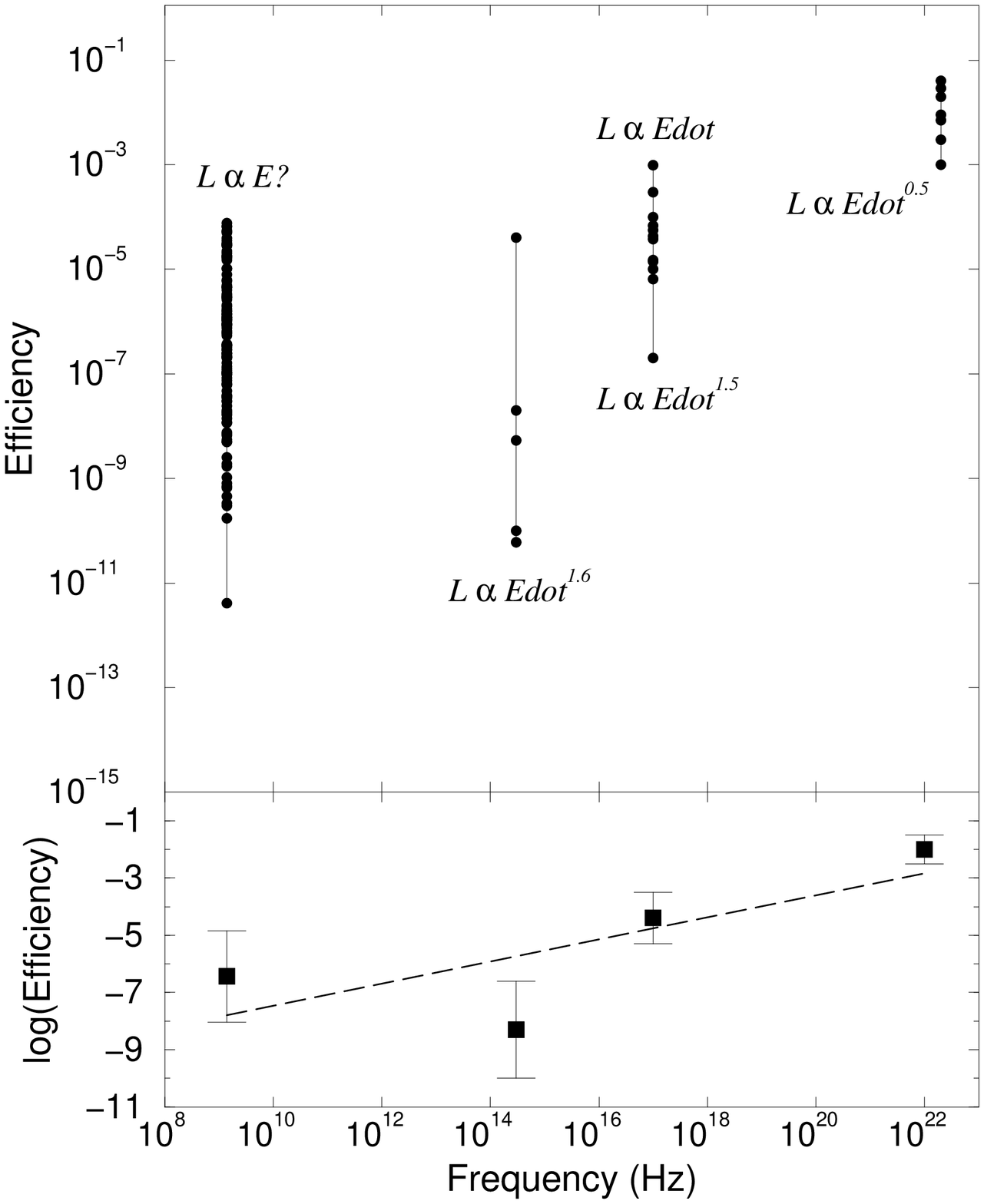,width=5.cm} &
\psfig{file=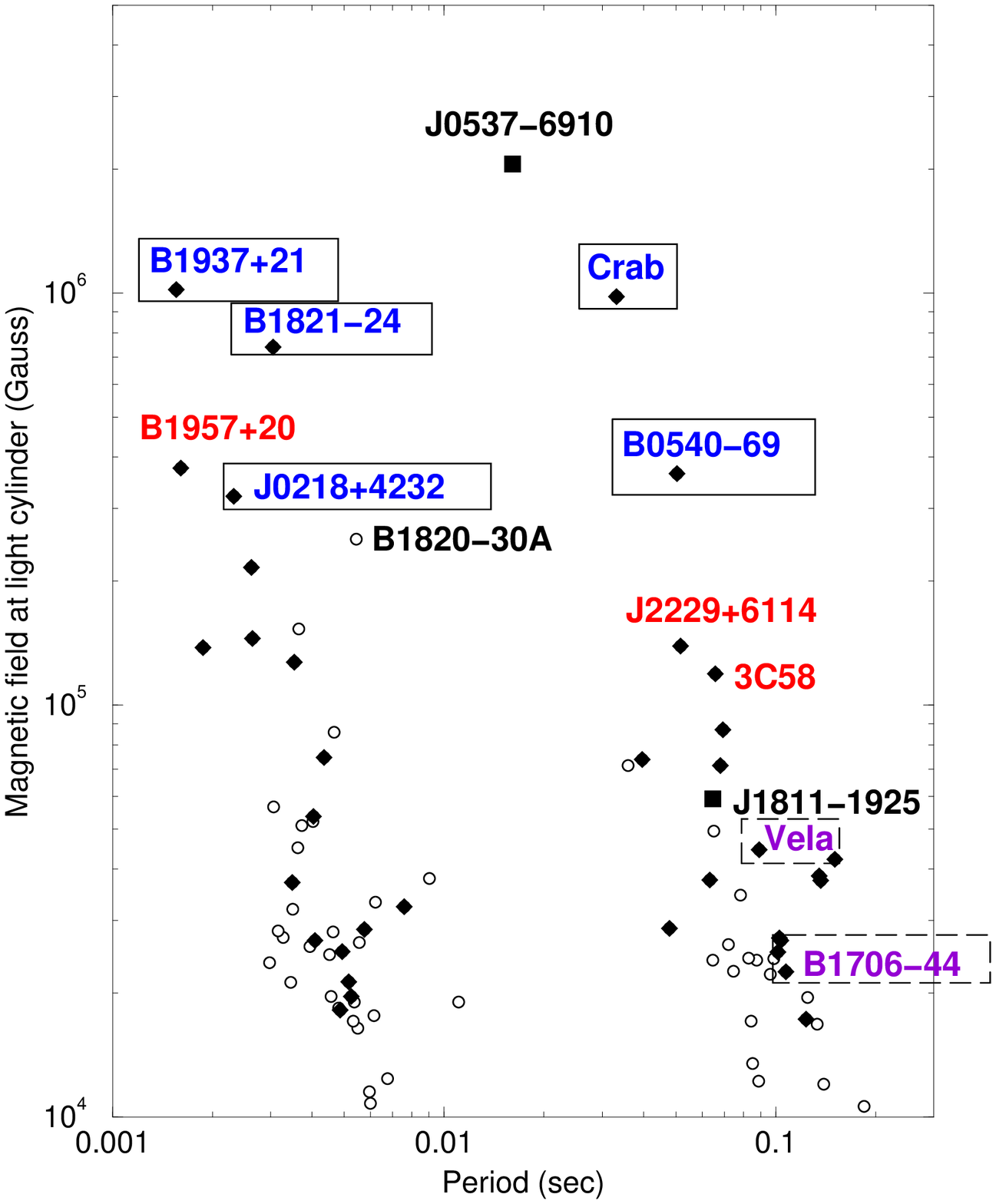,width=5.cm}
\end{tabular}
}

\caption{\label{fig:eff}
{\em Left:} Efficiency, $\eta\equiv L_\nu/\dot{E}$, as derived
for radio, optical, X-ray and $\gamma$-ray frequencies. A fit
to the median values shows an increase of efficiency with
frequency of $\eta \propto \nu^{0.17\pm 0.10}$. {\em Right:}
Magnetic field strength at the light cylinder. Filled circles
mark pulsars detected both at radio and high-energies. Radio-quiet
pulsars are shown as filled squares. Pulsars with detected giant
pulse emission are surrounded by a box. The Vela pulsar
and PSR B1706$-$44 show so-called ``micro-giants'' (see Johnston,
these proceedings.)
}
\end{figure}

\vspace{-0.5cm}

\section{Relation to spin-parameters}
\label{relation}

It would be naive to assume that the radiation processes are governed
by a single or a few control variables. It is nevertheless intriguing
to relate some of the radiation properties to the spin-down
luminosity and the magnetic fields at the surface and light cylinder.

A comparison of the energy output at radio, optical, X-ray and
$\gamma$-ray frequencies is complicated by uncertain distance
estimates, different spectral shapes, instrumental effects and in
particular by the unknown beaming fractions. The radio emission
appears to be emitted in a cone of radius $\rho$ that scales with
period as $\rho\propto 1/\sqrt{P}$ (e.g.~Kramer et al.~1998). Due to the
potential mixture of thermal and non-thermal emission observed, at
optical and X-rays one typically assumes isotropic emission in the
computation, even though this is obviously wrong for non-thermal
emission.  At $\gamma$-rays a beaming fraction of $1/4\pi$ is adopted
due to the lack of better knowledge. Bearing these uncertainties in
mind, we can try to compare the derived luminosities ``observed'' at
the different frequencies with the spin-down luminosity,
$\dot{E}$. At radio no correlation can be found between $L_{radio}$
and $\dot{E}$. This is not unexpected giving the importance of
geometry and the complicated processes that must be present to create
coherent emission. In the optical range, one can use peak luminosities
to mitigate geometrical issues or luminosities integrated over the
pulse profile (Shearer \& Golden 2002). Using the new distance for PSR
B0656+14 has a significant impact on the low number statistics, and we
find $L_{opt}\propto \dot{E}^{1.6\pm0.2}$ (see Fig.~\ref{fig:eff}). 
The X-ray luminosity,
$L_X$, was studied by a number of authors, including Becker \&
Tr\"u{}mper (1997) who found $L_X\propto \dot{E}$ at soft
X-rays ($0.1-2.4$ keV). More recently, Possenti et al.~(2002)
reviewed the harder X-rays ($2-10$ keV), deriving a relationship that
is remarkably similar to that at optical frequencies, $L_X \propto
\dot{E}^{1.5}$. At $\gamma$-rays, the low number statistics is again
affected by revised distances for PSRs B1055$-$52 and
J0218+4232 (Kramer et al.~2003a) but the values are consistent with 
$L_\gamma \propto \dot{E}^{0.5}$ (Thompson 2001). 
In conclusion, the inferred efficiencies, $\eta_\nu = L_\nu/\dot{E}$,
show large variation at each frequency band, but as a general trend
they increase from radio to $\gamma$-rays. A formal fit gives
$\eta_\nu \propto \nu^{0.17\pm 0.10}$ (Fig.~\ref{fig:eff}).

\label{giants}

In order to explain the lack of radio emission from magnetars (see
Kaspi, these proceedings), it has been suggested that a surface
magnetic field exceeding the quantum critical field, $B_{crit} =
m_e^2c^3/e\hbar = 4.4\times 10^{13}$ Gauss, would quench the radio
emission due to the lack of emitting plasma (Zhang \& Harding 2000,
see Fig.~\ref{fig:ppdot}).  However, the discoveries of pulsars with
magnetar-like spin-parameters and derived magnetic fields above the
critical field, like the 6-s PSR J1847$-$0130 with $B_S\sim 10^{14}$
Gauss (McLaughlin et al.~2003), question the importance of the actual
field value for the emission process. A more directly observable
effect may be caused by the magnetic field at the light cylinder. It
has been noted that pulsars with detected giant pulses happen to have
the largest field strengths (Fig.~\ref{fig:eff},
see contributions by Johnston and Joshi
et al.)  and that in general the giant pulse emission seems to align
with the high energy emission.  Indeed, there also appears to be a
direct observational link between the radio giant pulses of the Crab
and its optical emission (Shearer et al.~2003).  Giant pulses
may finally provide the clue to connect the emission across the whole
electromagnetic spectrum.

\vspace{-0.3cm}

\section{Summary \& Conclusions}

Apparently, the radio emission originates from close to the stellar
surface, while the high energy emission may tend to be created
further out. In that picture, one would not expect an alignment of
radio and high-energy emission, as typically observed. Where alignment
is observed, like for the Crab pulsar, the observed radio emission may
be of different origin and more related to that at high energies. Only
the Crab's pre-cursor component may be considered as the classical
radio pulse, while main and interpulse are by-products of high energy
processes causing also the High-Frequency Components at a few GHz.  If
that is the case, the Crab pulsar should be considered as a much less
luminous radio source, similar to PSR B0540$-$69 when ignoring its
giant pulse emission (Johnston \& Romani 2003).  In summary, we have
reason to believe that optical, X-ray and $\gamma$-ray processes
are related and that they connect to the radio via giant pulses.  A
lot appears to be determined, or at least influenced, by geometry, and
it is clear that no single control parameter exists. There is
still a lot to be done and understood.

\acknowledgments 
I thank the RAS for a substantial travel grant.


\end{document}